\documentclass[final,3p,10pt,times,twocolumn,sort&compress]{elsarticle}

\usepackage{amssymb}
\usepackage{gensymb}
\usepackage{booktabs}
\usepackage{multirow}
\usepackage{chemformula}

\newcommand{\BaLn}{Ba$_3$\textit{Ln}(BO$_3$)$_3$}
\newcommand{\hexsg}{$P6_3cm$}
\newcommand{\trigsg}{$R\bar{3}$}

\journal{Journal of Solid State Chemistry}

\begin{document}
\begin{frontmatter}

\title{Structure and magnetism of a new hexagonal polymorph of \ch{Ba3Tb(BO3)3} with a quasi-2D triangular lattice}

\author{Nicola D.~Kelly}
\author{Cheng Liu}
\author{Si\^{a}n E.~Dutton\corref{cor1}}
\ead{sed33@cam.ac.uk}
\cortext[cor1]{Corresponding author}
\address{Cavendish Laboratory, University of Cambridge, J J Thomson Avenue, Cambridge, CB3 0HE, UK}

\begin{abstract}
This article reports the structural and magnetic characterisation of a new hexagonal (\hexsg) low-temperature phase of \ch{Ba3Tb(BO3)3}, isostructural with the heavier lanthanide borates with formula \BaLn\ (\textit{Ln}~= Dy--Lu). The crystal structure contains a quasi-two-dimensional (2D) triangular lattice of Tb$^{3+}$ ions and is predicted to display unusual magnetic behaviour as a result of the low dimensionality. Magnetic susceptibility $\chi(T)$ shows antiferromagnetic interactions ($\theta_{CW} = -7.15$~K) and no sharp ordering transition at $T \geq 2$~K, but analysis of $\chi'(T)$, isothermal magnetisation and heat capacity suggests the possibility of short-range ordering at $T \approx 10$~K. The material exhibits an inverse magnetocaloric effect at $T < 4$~K.
\end{abstract}


%
\end{frontmatter}

\section{Introduction}
\begin{figure}[htbp] 
\centering
\includegraphics[width=6cm]{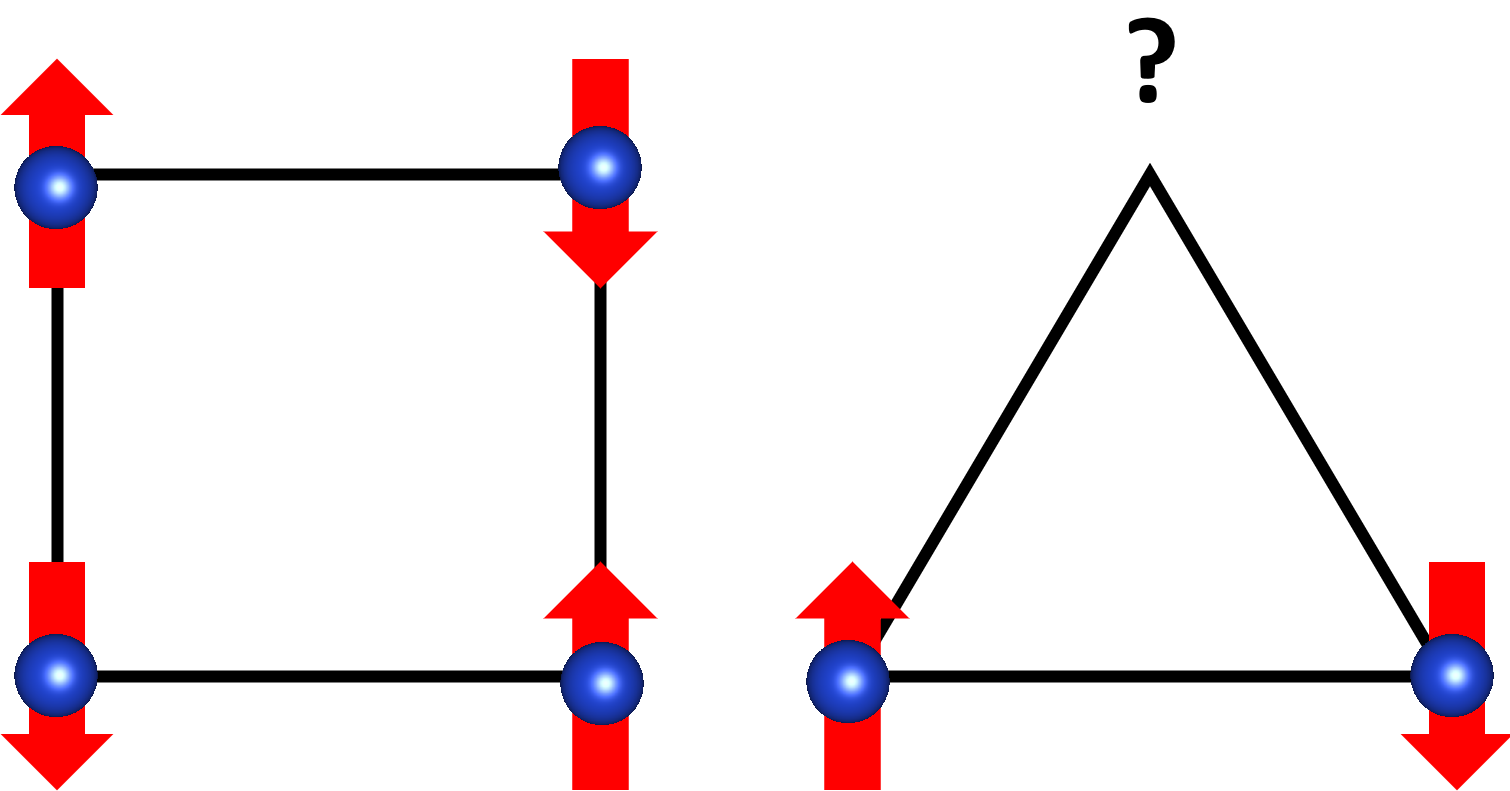}
\caption{Geometric frustration of magnetic spins on square and triangular plaquettes.}
\label{fig:frustration}
\end{figure}

A unique ordered ground state is possible on a square plaquette of magnetic spins with antiferromagnetic nearest-neighbour interactions, but not on a triangular plaquette (Fig.~\ref{fig:frustration}). This idea of geometric `frustration' can be extended to cover lattices in two or three dimensions (2D/3D) with a large number of spins, including more complex arrangements such as the kagome, garnet and pyrochlore lattices \cite{Ramirez1994,Collins1997}. From this simple observation comes a vast body of theoretical and experimental work on the manifestation of these lattices in real materials.

Frustration leads to degeneracy of the ground state and lowering or complete suppression of the magnetic ordering transition temperature. Some geometrically frustrated magnets (GFMs) have been proposed as quantum spin liquids (QSLs): states where the spins are highly entangled over long distances, but which do not exhibit either magnetic ordering or spin-freezing down to the lowest measured (millikelvin) temperatures \cite{Banerjee2016,Knolle2019}. The quasi-2D triangular QSL candidate \ch{YbMgGaO4} is a well-known example \cite{Knolle2019,Li2015,Rau2018,Li2020}. There is no magnetic site mixing, producing a structurally perfect quasi-2D triangular lattice which does not order down to 60~mK \cite{Li2015}. There is, however, random site disorder between Mg$^{2+}$ and Ga$^{3+}$ in the non-magnetic layers, which is expected to affect the crystal electric field environment of the Yb$^{3+}$ ions by distorting the local Yb--O geometry. A number of experimental studies have postulated that this factor is crucial in producing the QSL phase \cite{Paddison2017,Li2017}, while others predict that it should lead to spin-glass behaviour rather than a QSL \cite{Rau2018,Zhu2017,Ma2018}. Further examples include \ch{PrMgAl11O19}, \ch{NdMgAl11O19}, KBa\textit{Ln}(BO$_3$)$_2$ and RbBa\textit{Ln}(BO$_3$)$_2$, which feature perfect triangular lattices of magnetic ions but again suffer from non-magnetic site disorder \cite{Ashtar2019,Gao2011,Sanders2017,Guo2019}, and the completely disorder-free \ch{NaYbO2} which is a promising QSL candidate \cite{Ding2019,Bordelon2019}. In other materials the triangular lattice is distorted from perfect equilateral geometry, which may lead to commensurate or incommensurate magnetic ordering as the lattice is less frustrated \cite{Smirnov2009,Fishman2010,Dutton2011,Chapon2011a,Clark2019}.

\begin{figure*}[htbp] 
\centering
\includegraphics[width=14cm]{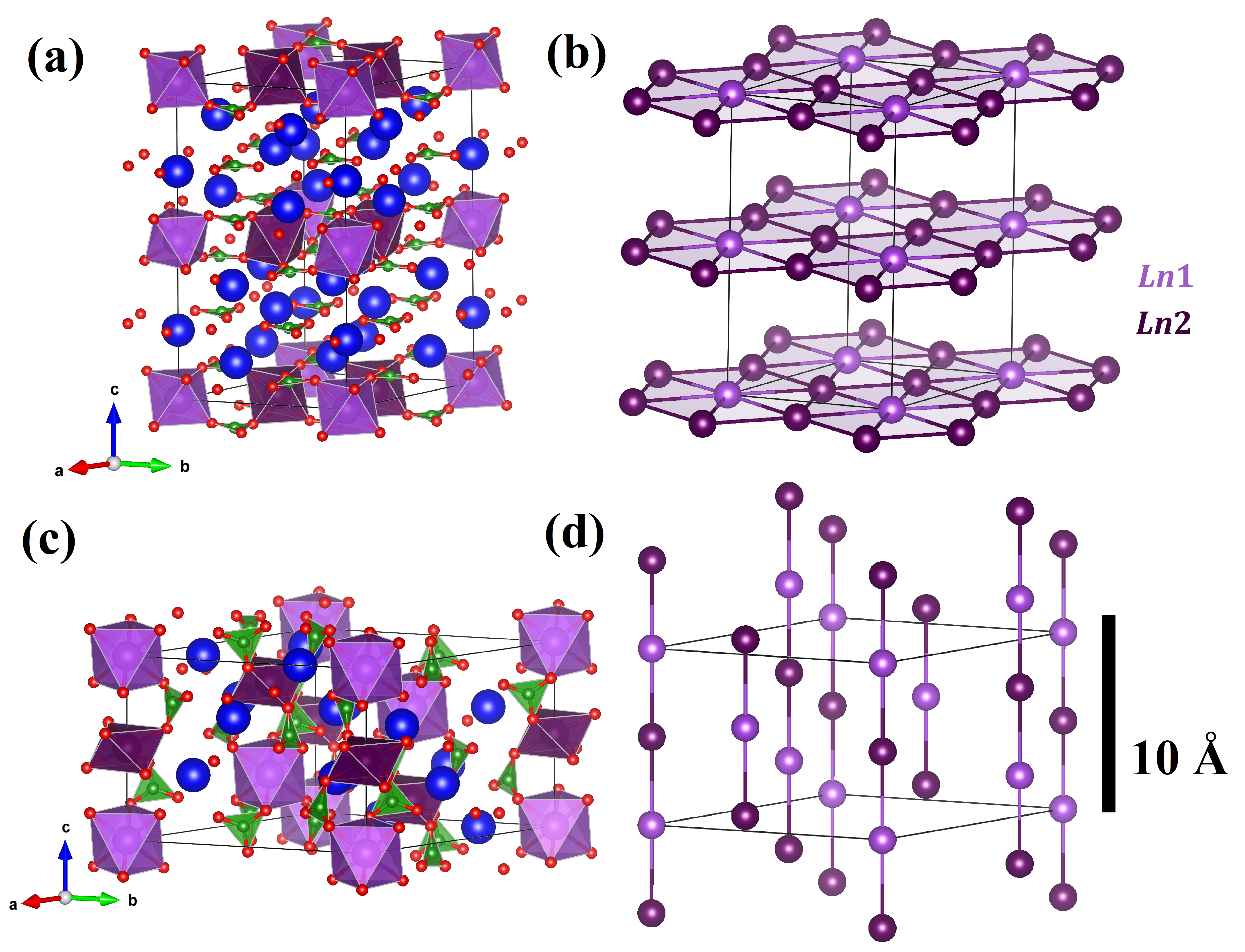}
\caption{Polyhedral models and magnetic lattices of the hexagonal \hexsg\ (a,b) and trigonal \trigsg\ (c,d) polymorphs of \BaLn. \textit{Ln} atoms and polyhedra in purple, Ba atoms in blue, B atoms in green and O atoms in red. All structures are shown on the same scale.}
\label{fig:balnstructurecomp}
\end{figure*}

Layered borates with the general formula \BaLn\ were first reported in 1993 by Ilyukhin and Dzhurinsky for \textit{Ln}~= Dy, Ho, Tm, Yb, Lu, with two different structures: a trigonal unit cell for \ch{Ba3Dy(BO3)3} and a hexagonal unit cell for the samples with \textit{Ln}~= Ho--Lu \cite{Ilyukhin1993}. These two crystal structures are shown in Fig.~\ref{fig:balnstructurecomp}, highlighting the arrangement of the lanthanide ions in each case. Both polymorphs of \BaLn\ contain triangular BO$_3^{3-}$ groups and 9-coordinate Ba$^{2+}$ ions. In the trigonal polymorph the lanthanide ions are arranged in closely spaced chains parallel to the $c$-axis, with adjacent chains slightly offset along that direction, while in the hexagonal polymorph the lanthanide ions form an eclipsed, quasi-2D triangular lattice parallel to the $ab$ plane. The compounds Sr$_3$\textit{Ln}(BO$_3$)$_3$ also crystallise in the trigonal structure type \cite{Schaffers1994}. A later study \cite{Khamaganova1999} found \ch{Ba3Dy(BO3)3} to have the hexagonal crystal structure, and in 2018, magnetic susceptibility measurements of the isostructural series \BaLn\ (\textit{Ln}~= Dy--Yb) were reported \cite{Gao2018}. This revealed antiferromagnetic nearest-neighbour interactions with neither long-range order nor glassiness at $T\geq2$~K, indicating that the compounds are geometrically frustrated and could be considered QSL candidates. A further study on \ch{Ba3Yb(BO3)3} used nuclear magnetic resonance (NMR) spectroscopy to show an absence of spin ordering or freezing down to $T=0.26$~K, including in high fields up to 15.8~T \cite{Zeng2019a}.

This article reports the synthesis, structural analysis and magnetic characterisation of a new hexagonal polymorph of \ch{Ba3Tb(BO3)3}, shown to be isostructural with the \BaLn\ analogues where \textit{Ln}~= Dy--Lu. Like the latter compounds, the new phase reported here displays no sharp magnetic ordering transition above 2~K, consistent with low-dimensional behaviour.

\section{Experimental}
A polycrystalline sample of hexagonal \ch{Ba3Tb(BO3)3} was synthesised according to a ceramic procedure, adapted from ref.~\citenum{Gao2018}, from \ch{BaCO3} (Alfa Aesar, 99.997~\%), \ch{H3BO3} (Alfa Aesar, 99.999~\%) and \ch{Tb4O7} (Alfa Aesar, 99.998~\%). Stoichiometric amounts of the reagents were ground with a pestle and mortar and placed in an alumina crucible. The powder was first heated in air at 500~\degree C for 12 h in order to decompose \ch{BaCO3} and \ch{H3BO3}. The sample was subsequently cooled, reground, pressed into a 13 mm pellet and reheated in air to 875~\degree C for several days with regrinding every 12 hours. Samples of \BaLn\ (\textit{Ln}~= Gd, Dy, Ho, Er, Y) were prepared in the same way from the relevant rare earth oxides and heated for 24~h at 900~\degree C. Each sample was then repelletised and reheated several times with the reaction temperature increasing in 50~\degree C increments up to 1250~\degree C.

Room temperature powder X-ray diffraction (PXRD) patterns were collected on a Bruker D8 diffractometer (Cu K$\alpha$, $\lambda = 1.541$ \AA) in the range $10 \leq 2\theta(\degree) \leq 90$ with a step size of 0.01\degree, 1 second per step. Low-temperature measurements, $12 \leq T$(K) $\leq 300$, were taken on cooling with the same diffractometer using an Oxford Cryosystems PheniX stage. The sample stage was allowed to thermalise at each temperature for 20 minutes before measuring across the range $10 \leq 2\theta(\degree) \leq 70$, step size 0.02\degree, 0.6 seconds per step. Rietveld refinement \cite{Rietveld1969} was carried out using the program TOPAS \cite{Coelho2018}.

Magnetic susceptibility was measured on a Quantum Design MPMS 3 in the temperature range $2 \leq T$(K) $\leq 300$. Isothermal magnetisation was measured on a Quantum Design PPMS Dynacool using the ACMS-II option in the field range $0 \leq \mu_0H$(T) $\leq 14$. In a low field of 500~Oe, the $M(H)$ curve is linear for all $T$ and the susceptibility can therefore be approximated by $\chi(T)=M/H$. Zero-field heat capacity was measured on the PPMS in the temperature range $2\leq T$(K) $\leq 30$. Samples were mixed with an equal mass of Ag powder (Alfa Aesar, 99.99~\%, --635 mesh) to improve thermal conductivity before being pressed into a 1~mm thick pellet for measurement. Apiezon N grease was used to provide thermal contact between the sample platform and the pellet. The contribution of Ag to the total heat capacity was subtracted using values from the literature \cite{Smith1995} and the \ch{Ba3Tb(BO3)3} lattice contribution was estimated and subtracted using a Debye model with $\theta_\mathrm{D}=255$~K \cite{Gopal1966}.

\section{Results and Discussion}
\subsection{Crystal Structure}
\begin{figure}[htbp] 
\centering
\includegraphics[width=7cm]{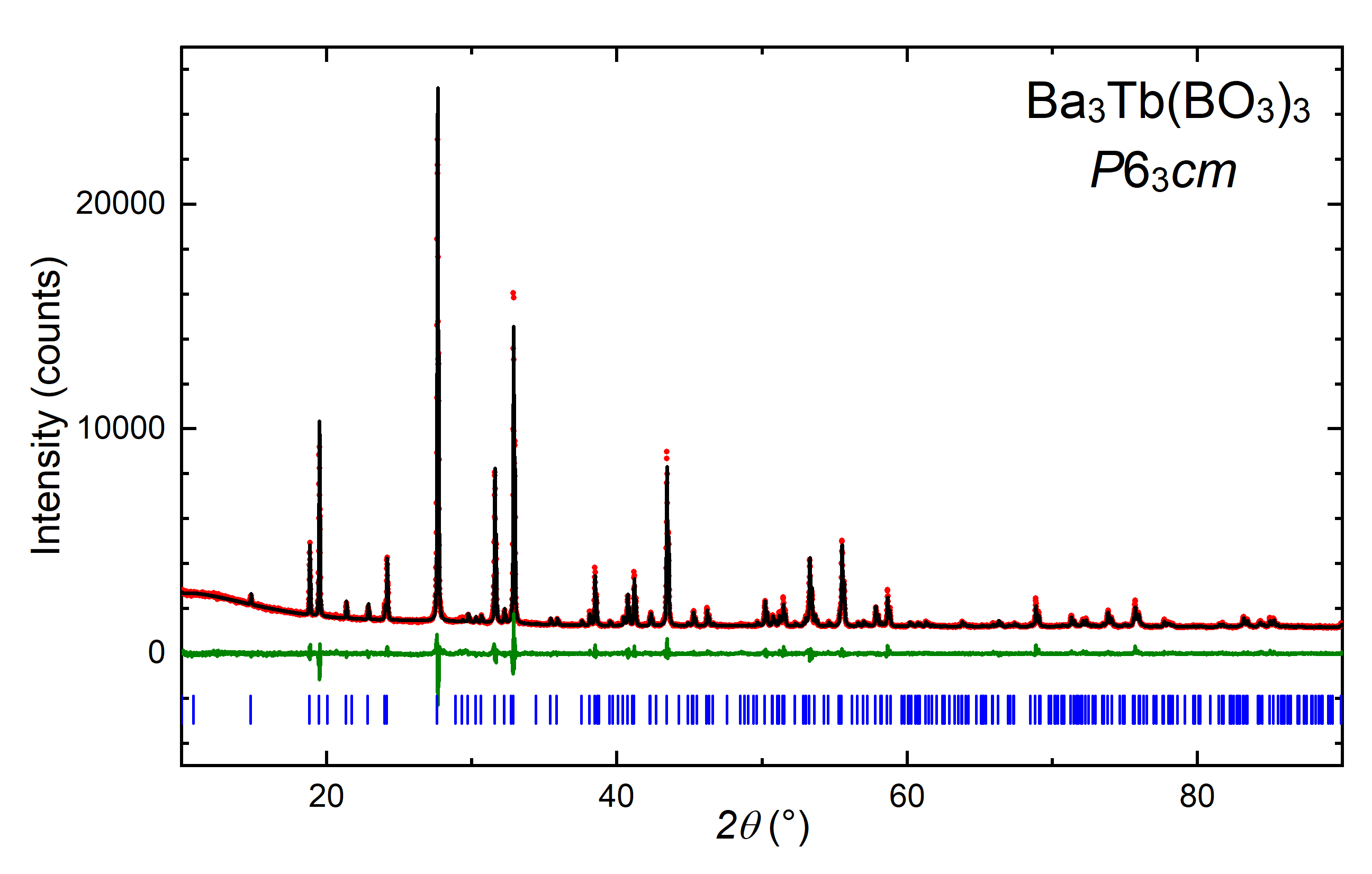}
\caption{Room temperature PXRD pattern for \ch{Ba3Tb(BO3)3} (\hexsg): Red dots -- experimental data; black line -- calculated intensities; green line -- difference pattern; blue tick marks -- Bragg reflection positions.}
\label{fig:rtpxrdbatbp}
\end{figure}

The structure of the new phase was investigated using PXRD (Fig.~\ref{fig:rtpxrdbatbp}; Table~\ref{table:batbatoms}) and found to be isostructural with the heavier \BaLn\ analogues, \textit{Ln}~= Dy--Lu (space group \hexsg); no reflections from the high-temperature \trigsg\ phase were observed. The heavier Ba and Tb atoms scatter X-rays much more strongly than B or O. In the refinement, therefore, the B and O atoms were fixed at the positions taken from the isostructural compound, \ch{Ba3Yb(BO3)3} \cite{Khamaganova1999}.

The presence, if any, of Ba$^{2+}$/Tb$^{3+}$ site disorder was tested by setting a suitable mixed cation occupancy in each of the six metal sites and refining partial occupancies with the overall stoichiometry fixed at 3:1. No significant site disorder was observed, in agreement with previous results for this structure type \cite{Ilyukhin1993,Khamaganova1999,Gao2018,Cox1994}. The result is as expected given the large difference in ionic radii between Ba$^{2+}$ (135 pm) and Tb$^{3+}$ (92.3~pm; each radius given for a 6-coordinate ion) \cite{Shannon1976}. The nearest-neighbour Tb--Tb distances (within the layers) are 5.45232(8)~\AA\ (Tb1--Tb2) and $\frac{a}{\sqrt{3}}=5.45193(4)$~\AA\ (Tb2--Tb2), while the perpendicular interlayer distance is $\frac{c}{2}=8.86021(8)$~\AA. The 2D layers are thus distorted from perfectly equilateral triangles, but the deviation in bond angle is smaller than 0.01~\%. This value is small compared with other 2D triangular lattice compounds such as the lanthanide orthoborates \textit{Ln}BO$_3$ (1.5~\%\ deviation) and RbBa\textit{Ln}(BO$_3$)$_2$ (2.5~\%\ deviation) \cite{Mukherjee2018a,Guo2019}.

\begin{table}[htbp] 
\centering
\caption{Atomic positions for \ch{Ba3Tb(BO3)3} (\hexsg). The refined lattice parameters are: $a$ = $b$ =  9.44302(6) \AA; $c$ = 17.72041(15) \AA; $V$ = 1368.440(20) \AA$^3$ ($Z$ = 6). The goodness-of-fit parameter $\chi^2 = 3.11$ and $R_{wp} = 7.47$. B and O atomic positions are taken from reports of the isostructural \ch{Ba3Yb(BO3)3} \cite{Khamaganova1999} and were not refined. $B_{iso}$ was fixed at 1 \AA$^2$ for all atoms.}
\label{table:batbatoms}
\begin{tabular}{c c c c c}
\toprule
Atom & Site & $x$ & $y$ & $z$\\
\midrule
Tb1 & $2a$ & 0 & 0 & 0 \\  
Tb2 & $4b$ & $\nicefrac{2}{3}$ & $\nicefrac{1}{3}$ & 0.9963(3) \\ 
Ba1 & $2a$ & 0 & 0 & 0.2262(5) \\  
Ba2 & $4b$ & $\nicefrac{2}{3}$ & $\nicefrac{1}{3}$ & 0.2711(4) \\  
Ba3 & $6c$ & 0.3409(13) & 0.3409(13) & 0.1307(6) \\  
Ba4 & $6c$ & 0.3227(13) & 0.3227(13) & 0.3682(5) \\  
B1 & $6c$ & 0.341 & 0.341 & 0.575 \\  
B2 & $6c$ & 0.345 & 0.345 & 0.746 \\  
B3 & $6c$ & 0.332 & 0.332 & 0.922 \\  
O1 & $6c$ & 0.187 & 0.187 & 0.584 \\
O2 & $12d$ & 0.328 & 0.478 & 0.574 \\
O3 & $12d$ & 0.185 & 0.336 & 0.746 \\
O4 & $6c$ & 0.483 & 0.483 & 0.747 \\
O5 & $12d$ & 0.338 & 0.476 & 0.921 \\
O6 & $6c$ & 0.196 & 0.196 & 0.932 \\
\bottomrule
\end{tabular}
\end{table}

Crystallographic studies in the 1990s, including both powder and single-crystal studies, found that \BaLn\ crystallises in the hexagonal space group \hexsg\ for the smaller ions, \textit{Ln}~= Dy, Ho, Y, Er, Tm, Yb, Lu and Sc, but the trigonal space group \trigsg\ for the larger \textit{Ln}~= Pr--Tb \cite{Khamaganova1999,Cox1994}. Analysis of the solid solution Ba$_3$Y$_{1-x}$Eu$_x$(BO$_3$)$_3$ gave a limit of $x=0.20(5)$ for incorporation of the larger Eu$^{3+}$ ion into the hexagonal structure. This limit corresponds to a maximum effective \textit{Ln}$^{3+}$ radius of 1.049(2)~\AA, consistent with the 6-coordinate Dy$^{3+}$ crystal radius of 1.052~\AA\ \cite{Shannon1976}. Our study finds that the maximum \textit{Ln}$^{3+}$ radius is greater than previously reported, at 1.063~\AA, the crystal radius of Tb$^{3+}$.

\begin{figure}[htbp] 
\centering
\includegraphics[width=7cm]{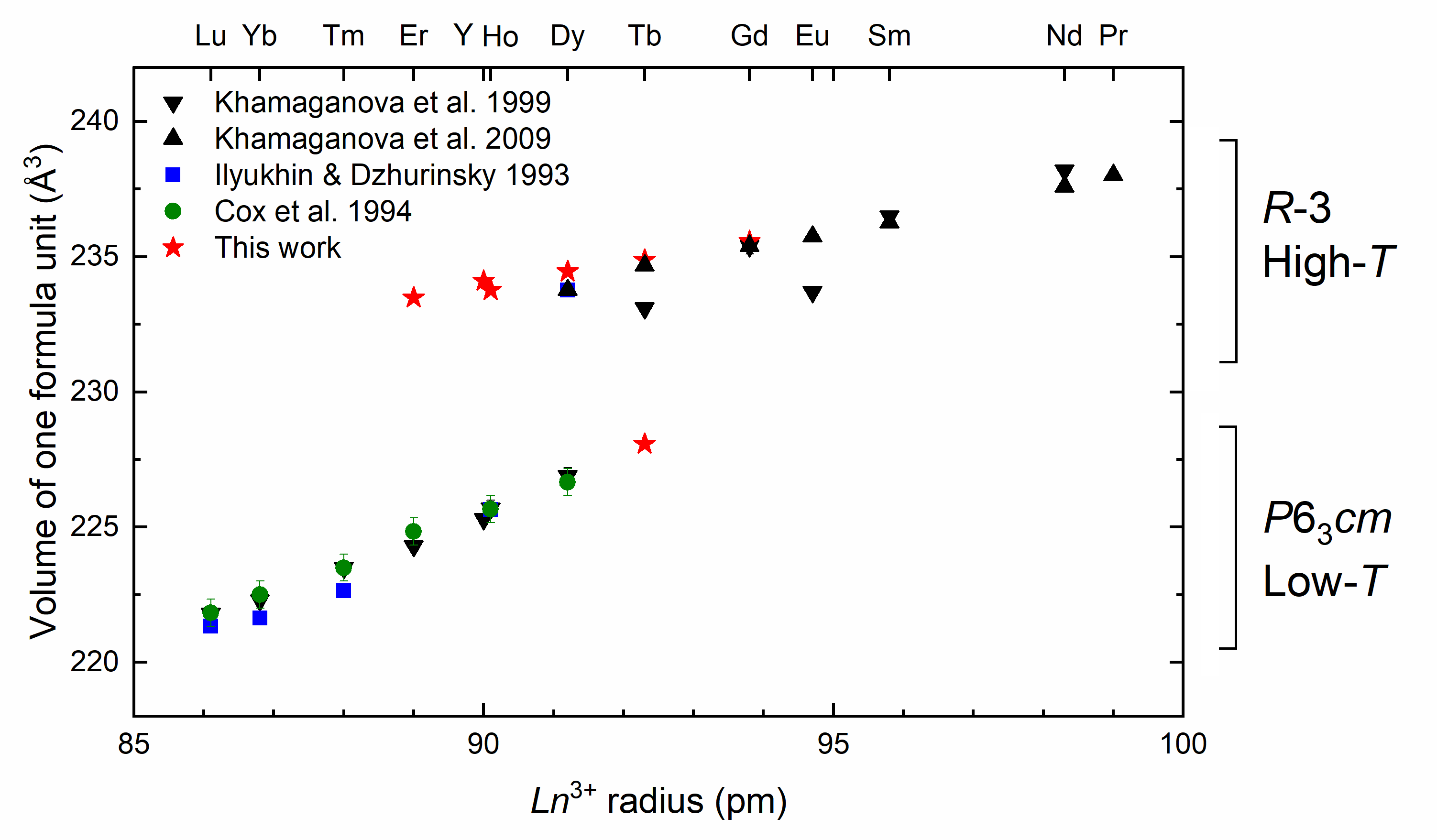}
\caption{Unit cell volumes of \BaLn\ compounds as a function of lanthanide ionic radius \cite{Ilyukhin1993,Cox1994,Khamaganova1999,Khamaganova2009}.}
\label{fig:balnvolumes}
\end{figure}

For certain compounds, particularly \ch{Ba3Dy(BO3)3}, there is debate in the literature over the melting point and whether or not the hexagonal phase transforms into the trigonal phase at high temperatures \cite{Khamaganova1999,Cox1994,Simura2017}. Fig.~\ref{fig:balnvolumes} summarises the structural parameters for all \BaLn\ phases reported so far. The existence of both \trigsg\ and \hexsg\ polymorphs for \textit{Ln}~= Dy, Tb, depending on the synthesis conditions, raised the possibility of other polymorphs of \BaLn\ being achievable. We observed full conversion from the hexagonal to the trigonal structure for \textit{Ln} = Dy, Ho and Er (but not Yb or Lu) when heated at higher temperatures in the range 1050--1250~\degree C, as found by Simura \textit{et al}. \cite{Simura2017}. It is plausible that synthesis at lower temperatures might produce hexagonal \ch{Ba3Gd(BO3)3}, but this is unlikely to be the case for even lighter \textit{Ln}, as a larger \textit{Ln}:\textit{A} radius ratio (\textit{A}~= Ba, Sr) favours the trigonal structure: the compounds with \textit{A}~= Sr exhibit only the trigonal phase \cite{Schaffers1994}.

The low-temperature phase of \ch{Ba3Tb(BO3)3} is stable when heated from room temperature up to the synthesis temperature (875~\degree C) for prolonged periods of time, e.g.~1 week or more. At 875--900~\degree C the high-temperature (\trigsg) phase is the initial synthesis product, but prolonged heating gradually converts it to the low-temperature (\hexsg) phase. This reaction took about 360 hours, i.e.~15~x~24 hours with intermediate regrinding (see Supplementary Information, Fig.~S1). The process is reversible: prolonged heating at $T\geq1000$~\degree C produces the trigonal (high-temperature) phase once more.

\begin{figure}[htbp] 
\centering
\includegraphics[width=7cm]{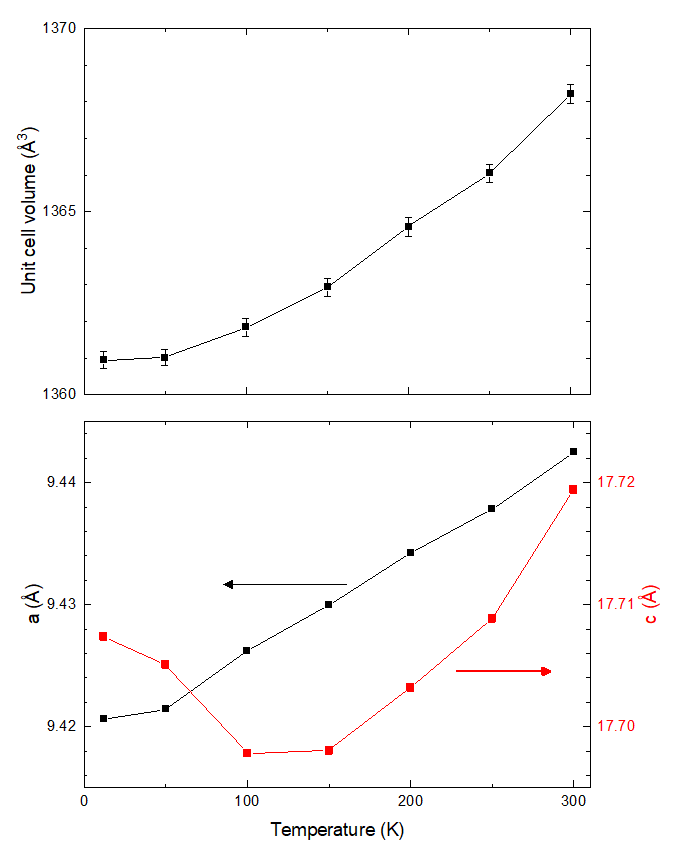}
\caption{Lattice parameters of \ch{Ba3Tb(BO3)3} (\hexsg), from PXRD, as a function of temperature. For $a$ and $c$ some error bars are smaller than the datapoints. Solid lines are a guide to the eye.}
\label{fig:vtparams}
\end{figure}

Variable-temperature PXRD was carried out on the new low-temperature phase of \ch{Ba3Tb(BO3)3}. No structural phase transition was observed on cooling from 300 to 12 K. The refined lattice parameters are given in Fig.~\ref{fig:vtparams}. The $a$ lattice parameter shows a monotonic decrease on cooling but the $c$ parameter (interplanar spacing) first decreases on cooling to 100~K, followed by a steady increase on further cooling. No low-temperature measurements on other \BaLn\ polymorphs have been reported, but similar behaviour has been observed in the layered compound $\alpha$-\ch{SrCr2O4} and in Ruddlesden-Popper oxides; in both cases this is associated with distortions of the layers \cite{Dutton2011,Chen2019a}. Rietveld refinements at each temperature showed no significant changes in the Tb atomic positions across the range 12--300~K, indicating that the magnetic layers retain their 2D triangular geometry, but there was a steady decrease in the $z$-position of the Ba1 site by 4~\%\ on cooling from 300~K to 12~K (see Supplementary Information, Fig.~S2). The changes in lattice parameters may also be related to changes in the B and O atomic positions, which cannot be observed directly using XRD; variable-temperature neutron diffraction studies are required to investigate this further. It is also plausible that this structural behaviour may be related to competition between polymorphs which coexist at a local level in the same sample \cite{Morelock2013,Baise2018}; in this case diffuse scattering measurements would be the key to solving the problem.

\subsection{Bulk magnetic properties}
\begin{figure}[htbp] 
\centering
\includegraphics[width=7cm]{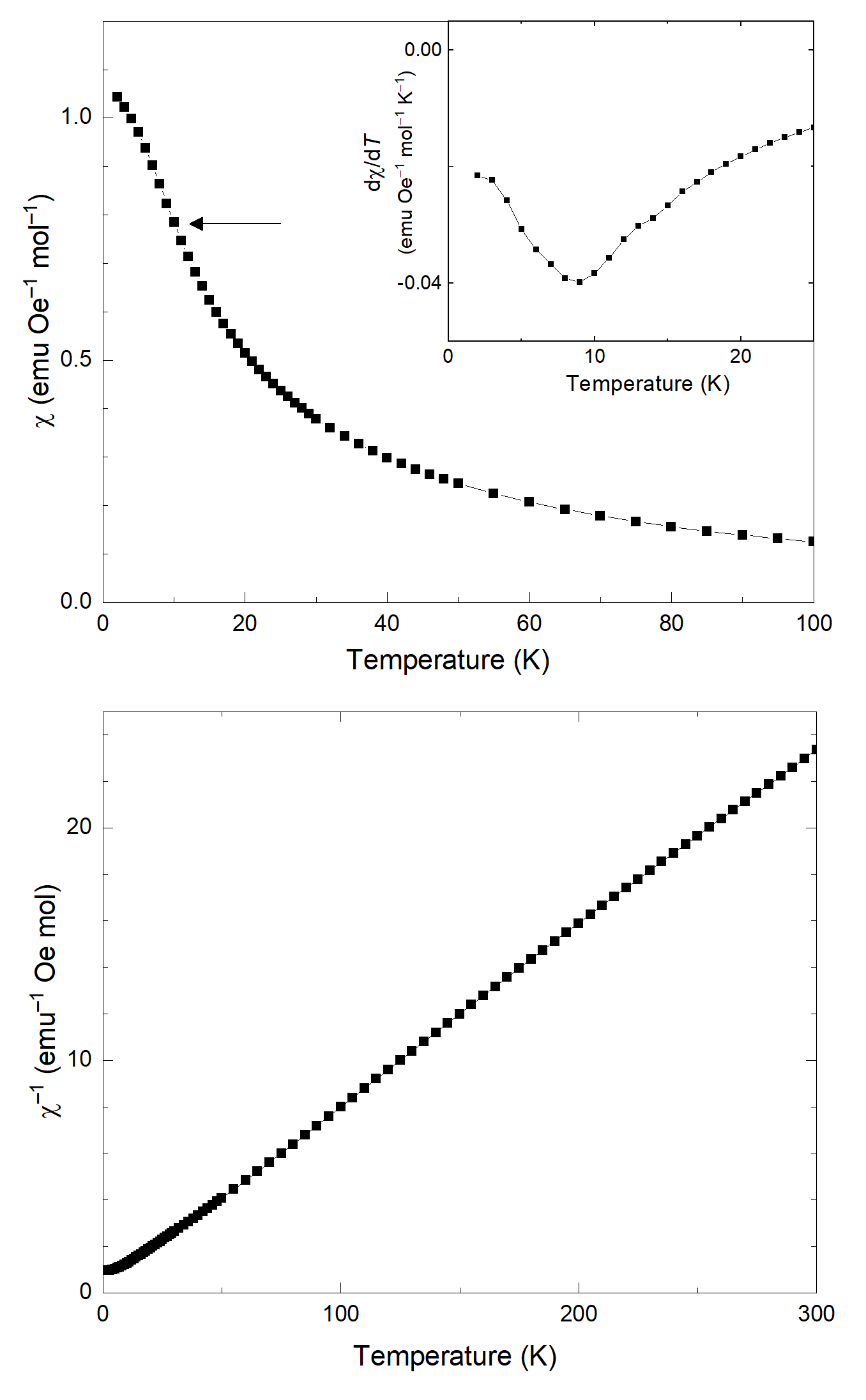}
\caption{Top: Magnetic susceptibility as a function of temperature for the new hexagonal phase of \ch{Ba3Tb(BO3)3} (inset: $d\chi/dT$ at low temperatures). The arrow indicates the location of the minimum in $d\chi/dT$ at $T=9$~K. Bottom: Reciprocal magnetic susceptibility, $\chi^{-1}$.}
\label{fig:batbmvtpanels}
\end{figure}

The DC magnetisation was measured on warming from 2 to 300~K in a field of 500~Oe after cooling the sample from 300~K in zero magnetic field (ZFC) (Fig.~\ref{fig:batbmvtpanels}) using a Quantum Design MPMS 3. Linear Curie-Weiss fitting, $\chi = C/(T-\theta_{CW})$, was carried out in both high-temperature (150--300~K) and low-temperature (20--50~K) regimes, in order to account for thermal population of excited states at higher temperatures \cite{Guo2019}. The resulting magnetic parameters are given in Table~\ref{table:batbmagdata}. The effective magnetic moment per Tb$^{3+}$ ion was calculated from the Curie constant using $\mu_{eff}(\mu_B) = \sqrt{8C}$ for both fitting regimes and agrees with the theoretical free-ion value $g_J\sqrt{J(J+1)}$ within experimental error. The Curie-Weiss temperature $\theta_{CW}$ is negative, indicating antiferromagnetic interactions. Geometric frustration was quantified using the frustration parameter $f=|\theta_{CW}/T_0|$, where $f>10$ usually indicates strong frustration \cite{Ramirez1994}, and the resultant ranges are given in Table~\ref{table:batbmagdata}. Further experiments are required in order to find $T_0$ and accurately assess the frustration of the magnetic lattice. Crystal field effects as well as short- and long-range dipolar interactions between the $^7$F$_6$ Tb$^{3+}$ ions are expected to contribute to the measured Curie-Weiss temperature \cite{Gingras2000}. Considering the relatively large distance between ions (5.45~\AA), the exchange interaction $J_{nn}$ is expected to be weak and cannot be obtained with confidence from the susceptibility data.

\begin{figure}[htbp] 
\centering
\includegraphics[width=7cm]{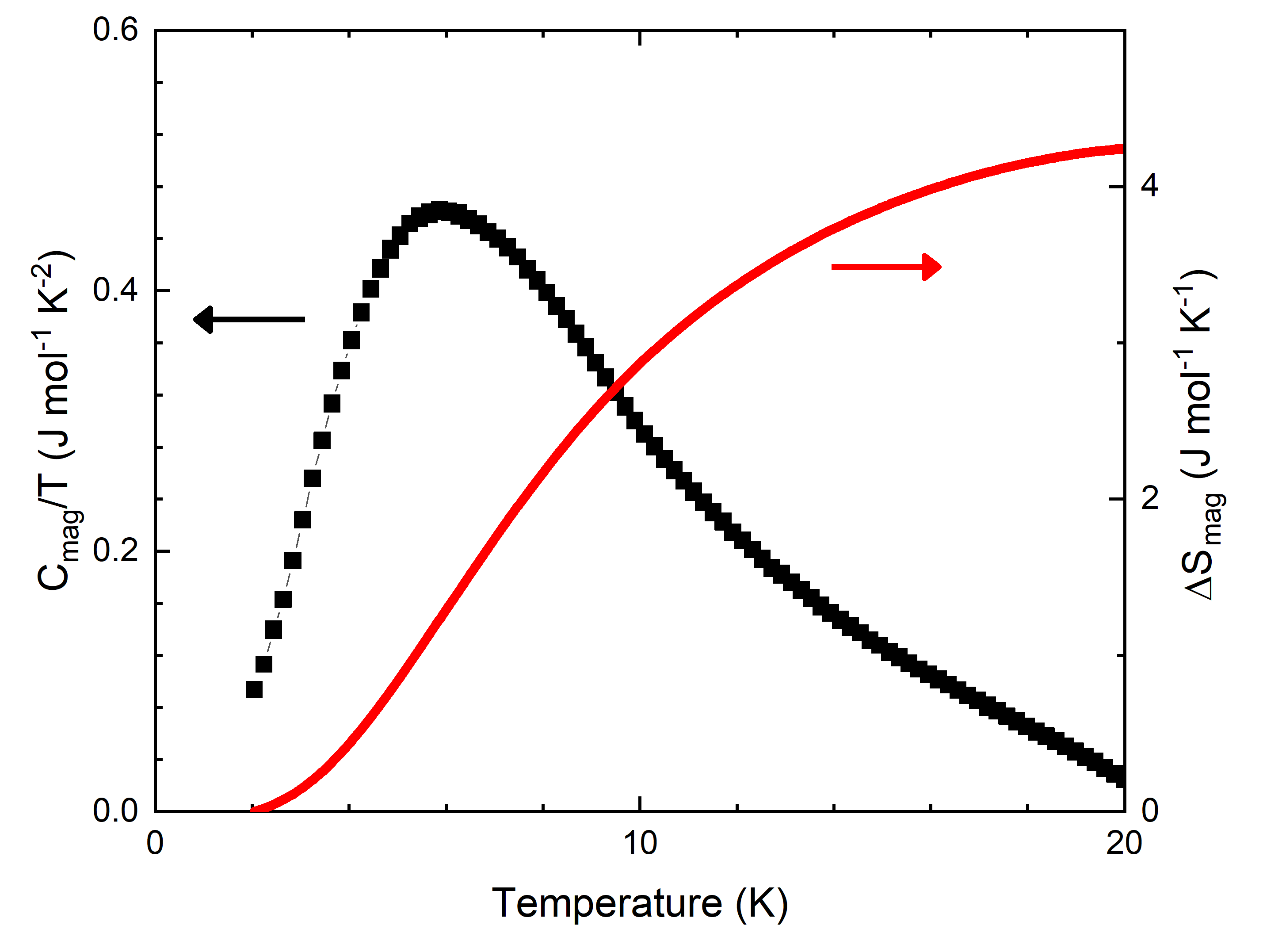}
\caption{Magnetic heat capacity $C_\mathrm{mag}/T$ and entropy $\Delta S$ of \ch{Ba3Tb(BO3)3} (\hexsg) in the range 2--20~K.}
\label{fig:batbhc}
\end{figure}

In the susceptibility curve, there is no sharp magnetic ordering transition at any temperature $T \geq$ 2~K. However, the change in gradient at $T<10$~K (Fig.~\ref{fig:batbmvtpanels}, inset) suggests the presence of a broad transition, e.g.~due to short-range ordering. In the tripod kagome lattice family \ch{\textit{A}2Tb3Sb3O14}, the compounds with \textit{A}~= Mg and Zn both show a similar broad feature in the specific heat capacity $C_{mag}$ around $1.5-2.5$~K, which the authors attribute to the nuclear spin degree of freedom of the Tb$^{3+}$ ion ($^{159}$Tb, $I = 3/2$), although no such peak is visible in the susceptibility curves \cite{Dun2017a}. The susceptibility feature in our measurements may alternatively arise from one of the two crystallographically unique Tb$^{3+}$ ions forming a singlet state. Such behaviour was observed in \ch{TbInO3}, which has a similar triangular lattice of inequivalent Tb$^{3+}$ ions; however, one-third of the ions in \ch{TbInO3} have a singlet ground state with a gap of $\Delta=7.5$~K to the first excited state, while the remaining two-thirds have a doublet ground state, producing an effective honeycomb lattice at $T<\Delta$ \cite{Clark2019}. The zero-field heat capacity of \ch{Ba3Tb(BO3)3} (Fig.~\ref{fig:batbhc}) shows a broad peak in $C_\mathrm{mag}/T$ at $T\approx 6$~K which is consistent with the hypothesis of short-range ordering in this temperature range. The magnetic entropy was estimated by integration of $C/T$ as 4.2~J~mol$^{-1}$~K$^{-1}$, which is just under 20~\%\ of the maximum magnetic entropy $R\ln(2J+1)=21.32$~J~mol$^{-1}$~K$^{-1}$. Neutron scattering and/or further magnetometry and heat capacity measurements on \ch{Ba3Tb(BO3)3} below 2~K are required in order to elucidate the exact reason for the observed behaviour. 

\begin{table}[htbp] 
\centering
\caption{Bulk magnetic properties of \ch{Ba3Tb(BO3)3} (\hexsg): f.u. = formula unit.}
\label{table:batbmagdata}
\begin{tabular}{c c c}
\toprule
 & Low-$T$ fit & High-$T$ fit \\
\midrule
$T_0$ (K) & $<2$ & $<2$ \\
$\theta_{CW}$ (K) & --7.15(14) & --9.8(2) \\
$f$ & $>3.5$ & $>4.9$ \\
Experimental $\mu_{eff}$ ($\mu_B$) & 10.6(2) & 10.3(2) \\
Theoretical $\mu_{eff}$ ($\mu_B$) & 9.72 & 9.72 \\
$g_JJ$ & 9 & 9\\
$M_{2K,14T}$ ($\mu_B$/f.u.) & 5.27(11) & 5.27(11) \\
\bottomrule
\end{tabular}
\end{table}

\begin{figure}[htbp] 
\centering
\includegraphics[width=7cm]{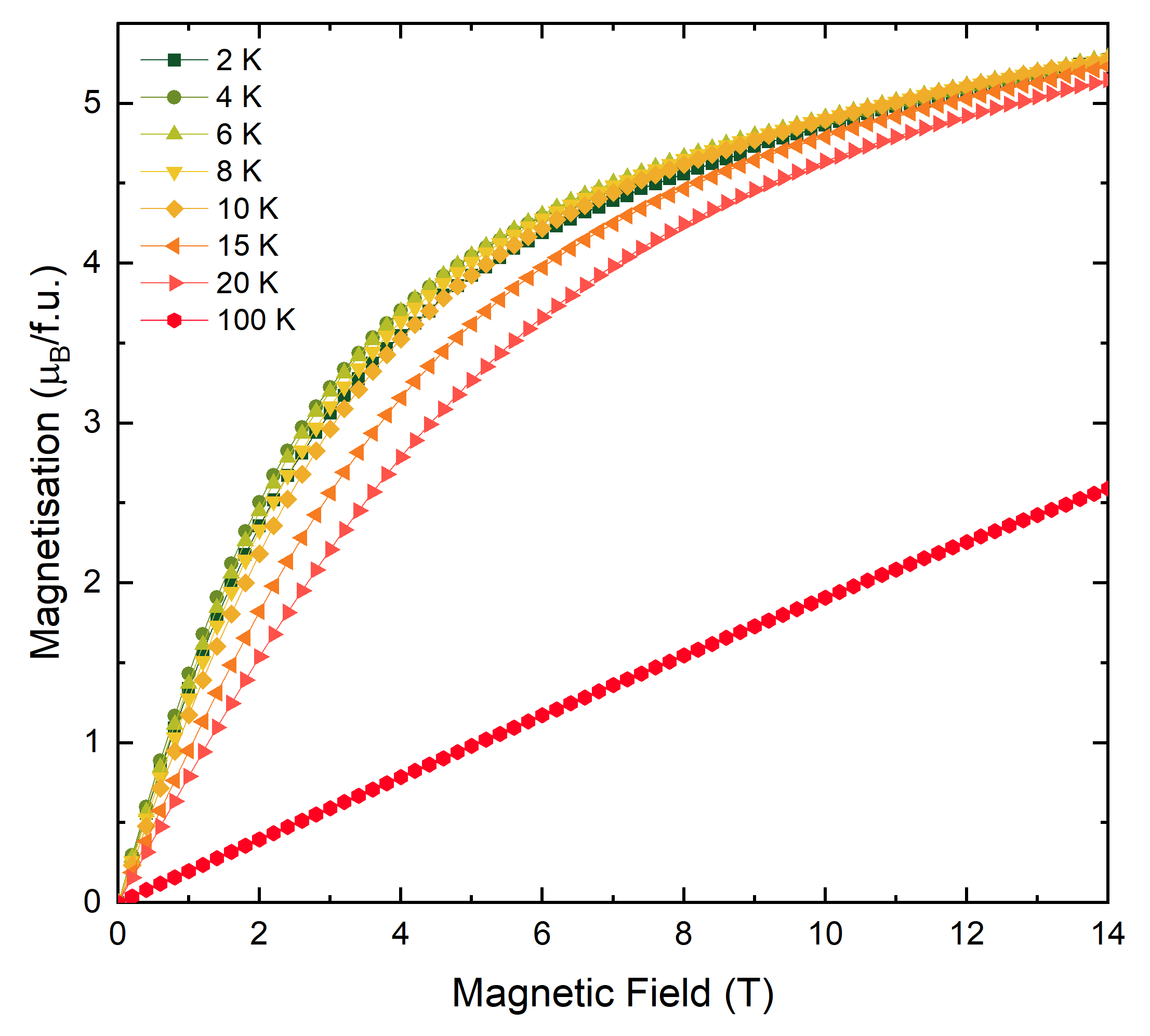}
\caption{Isothermal magnetisation curves for the new hexagonal phase of \ch{Ba3Tb(BO3)3}.}
\label{fig:batbmvh}
\end{figure}

Isothermal magnetisation was measured as a function of applied field using a Quantum Design PPMS (Fig.~\ref{fig:batbmvh}). At low temperatures, $T\leq20$~K, there is some curvature but no sign of saturation in fields up to 14 T. The magnetisation at 2~K and 14~T, 5.27~$\mu_B$ per formula unit (f.u.), is significantly lower than that expected for a fully spin-polarised sample, $g_JJ=9$~$\mu_B$/f.u., indicating strong single-ion anisotropy. The data for this new phase of \ch{Ba3Tb(BO3)3} strongly resemble recent measurements on \ch{KBaTb(BO3)2} and \ch{RbBaTb(BO3)2}, which also contain triangular lattices of Tb$^{3+}$ ions, and where the magnetisation also reaches approximately 4.5~$\mu_B$/f.u.~at 2~K, 9~T \cite{Sanders2017,Guo2019}. Magnetic measurements on a single crystal at different orientations to the applied field, along with CEF calculations from inelastic neutron scattering, would be required to confirm the local single-ion anisotropy of \ch{Ba3Tb(BO3)3}.

\subsection{Magnetocaloric effect}
The lowering of the magnetic ordering temperature due to frustration is exploited in magnetic refrigeration applications in order to reduce usage of liquid helium \cite{Pecharsky1999}. Such devices make use of the magnetocaloric effect (MCE) to lower the sample temperature by adiabatic demagnetisation. The strongest MCE is observed in compounds containing Gd$^{3+}$ due to its large effective magnetic moment (7.98~$\mu_B$) and weak crystal field interactions; the best known ceramic material is gadolinium gallium garnet (GGG) which consists of interlinked corner-sharing triangles of spins \cite{Barclay1982,Moessner2001}. In general the MCE is maximised in high fields ($\mu_0H > 5$~T), where Heisenberg systems have been shown to produce the greatest MCE. At low fields $\mu_0H \leq 2$~T, systems with significant single-ion anisotropy (usually containing lanthanides other than Gd) tend to perform better \citep{Saines2015}.

In order to calculate the MCE, isothermal magnetisation measurements were made at temperatures $T$ = 2, 4, 6, 8 and 10~K at field intervals of $\mu_0H=0.2$~T. The magnitude of the MCE for \ch{Ba3Tb(BO3)3} was quantified by calculating the change in magnetic entropy per mole, $\Delta S_m$, according to the Maxwell thermodynamic relation \citep{Pecharsky1999}:

\begin{equation} 
\Delta S_m = \int_{H_0}^{H_1} \left( \frac{\partial M}{\partial T}\right)_H dH
\label{equation:mce}
\end{equation}
before converting $\Delta S$ into gravimetric units (J~K$^{-1}$~kg$^{-1}$). While the molar MCE can be compared with the theoretical maximum $\Delta S_m$ for each lanthanide ion -- dependent on the electronic configuration -- the gravimetric MCE is more useful in comparing materials for practical application in devices.

The MCE is generally much weaker for Tb-containing compounds than their Gd analogues at equivalent fields and temperatures, due to the strong single-ion anisotropy of Tb$^{3+}$ compared with Gd$^{3+}$ \cite{Numazawa2003,Spaldin2011,Saines2015}. However, at low fields and moderate temperatures (4--10~K) some Tb-containing compounds have shown better magnetocaloric performance than their Gd analogues. Due in part to the large relative atomic mass of Ba, the maximum MCE of \ch{Ba3Tb(BO3)3} is very low compared with these benchmarks: in a field of 5~T, \ch{Ba3Tb(BO3)3} has $-\Delta S_{max}=1.22$~J~K$^{-1}$~kg$^{-1}$, while \ch{Tb(HCO2)3} and \ch{TbOHCO3} have 15.7 and 33.72~J~K$^{-1}$~kg$^{-1}$ respectively \cite{Saines2015,Dixey2018}.

\begin{figure}[htbp] 
\centering
\includegraphics[width=7cm]{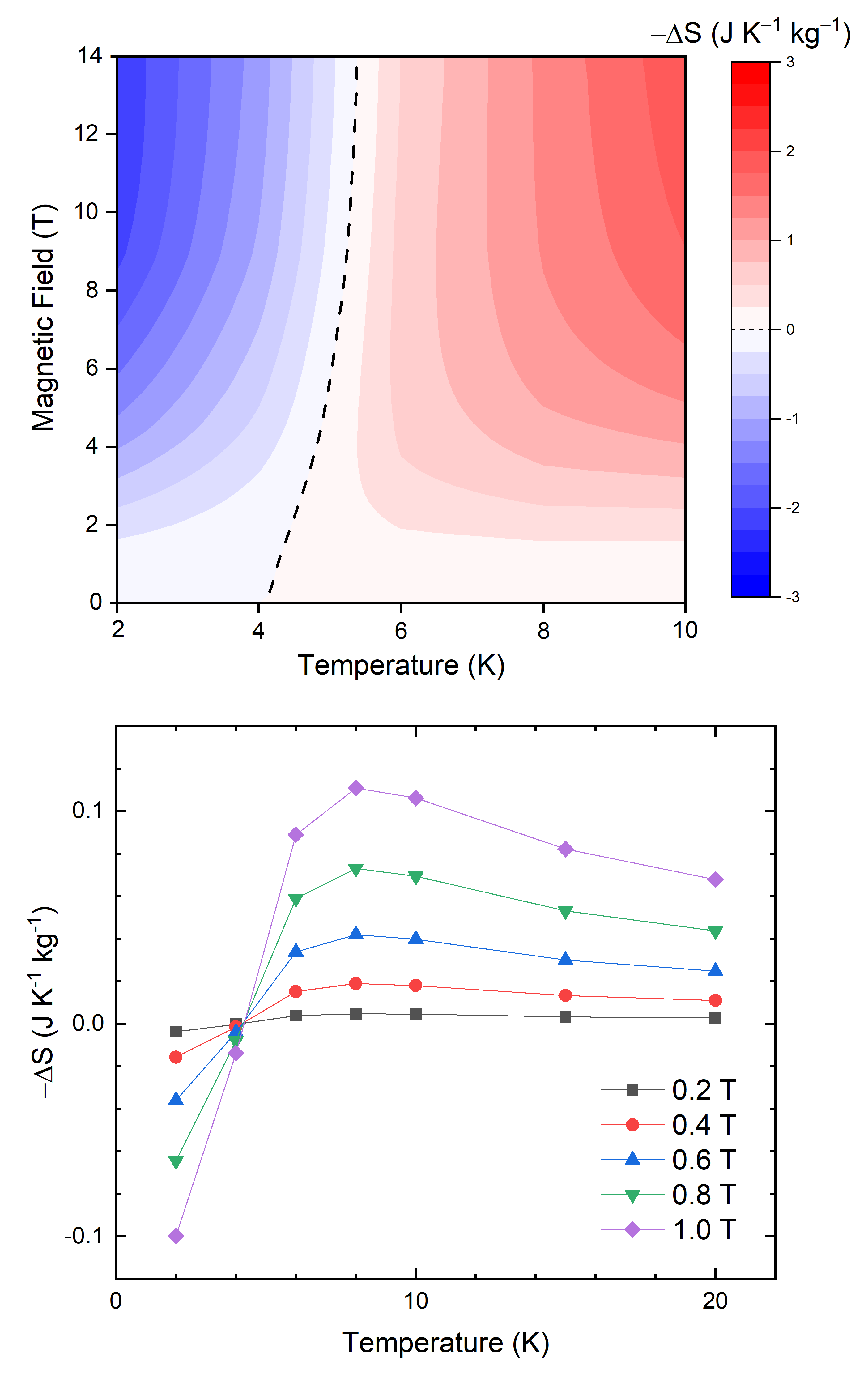}
\caption{Magnetocaloric data for the new hexagonal phase of \ch{Ba3Tb(BO3)3}. Top: $-\Delta S(H,T)$; the dashed line indicates $\Delta S=0$ (interpolated). Bottom: $-\Delta S(T)$ at fields $\mu_0H\leq1$~T.}
\label{fig:batbmcepanels}
\end{figure}

The magnetocaloric data are plotted as a function of field and temperature in Fig.~\ref{fig:batbmcepanels} using the convention of plotting $-\Delta S$, where a larger number indicates a stronger MCE. Data at intermediate temperature and field values were linearly interpolated.

The MCE shows features that are not expected for a material in a purely paramagnetic regime. At all fields $0 \leq \mu_0H$(T) $\leq 14$ a maximum in $-\Delta S$ is observed at $T\approx 8-9$~K, and on further cooling $\Delta S$ becomes positive. The maximum in $-\Delta S$ is at a similar temperature to the minimum in $d\chi/dT$ observed in the susceptibility measurement and may therefore indicate magnetic ordering, as observed in the substituted gallium garnets \textit{Ln}$_3$MnGa$_4$O$_{12}$ (\textit{Ln}~= Tb, Dy, Ho) \cite{Mukherjee2018}. Inverse MCE ($\Delta S>0$) is uncommon but has been observed due to crystal electric field effects in \ch{PrNi5} \cite{Pecharsky1999,VonRanke1998}, in transition metal alloys with first-order magnetic transitions \cite{Krenke2005}, and in uniaxial paramagnets containing non-Kramers ions (Pr, Tb, Ho, and Tm) \cite{Kokorina2017}. Susceptibility and specific heat measurements at fields $\geq500$~Oe and further $M(H)$ curves at smaller temperature intervals are planned in order to investigate this behaviour in \ch{Ba3Tb(BO3)3}.

\section{Conclusions}
We have synthesised a new hexagonal polymorph of \ch{Ba3Tb(BO3)3} and measured its bulk magnetic properties. The new phase is isostructural with the heavier analogues \BaLn\ (\textit{Ln}~= Dy--Lu) where the arrangement of the \textit{Ln}$^{3+}$ ions forms a quasi-2D triangular lattice. PXRD measurements confirm that there is no magnetic site disorder and that the layers of Tb$^{3+}$ ions are well separated. Magnetic measurements indicate that, like its heavier analogues, the title compound has antiferromagnetic interactions with no 3D magnetic ordering transition above 2~K, although there is evidence for possible short-range magnetic ordering and/or field-induced ordering in the $\chi'(T)$ and $M(H)$ data. Further experiments including measurements at $T<2$~K are required in order to determine the nature of this ordering and to investigate the magnetic ground state. In contrast to e.g.~\ch{YbMgGaO4}, the \textit{Ln}$^{3+}$ ions in \BaLn\ have well-defined environments as there is no non-magnetic site disorder.

Future work will investigate the possibility of forming an isostructural polymorph of \ch{Ba3Gd(BO3)3}, which may be of interest for magnetic refrigeration at liquid helium temperatures due to the large MCE associated with Gd$^{3+}$ ions. Further work will focus on the synthesis of \BaLn\ (\textit{Ln}~= Tb--Yb) at different temperatures in an attempt to resolve disputes in the literature regarding the high-$T$ phase behaviour.

\section{Acknowledgments}
We acknowledge funding from the EPSRC for a PhD studentship (NDK) and the use of the Advanced Materials Characterisation Suite (EPSRC Strategic Equipment Grant EP/M000524/1). NDK thanks P.~Mukherjee for useful discussions.

\section{Additional information}
The authors declare that there are no competing interests.

Supplementary information is available on request.

Data relating to this publication are available at https://doi.org/10.17863/CAM.55685

The structure reported in this manuscript has been deposited with the Cambridge Crystallographic Data Centre: CSD 2010478.

\bibliographystyle{elsarticle-num} 
\bibliography{library}

\end{document}